Managing Renewable Energy Resources Using Equity-Market Risk Tools - the Efficient Frontiers

Haim Grebel[1,3] Divya Vikas[2], Jim Shi[2,3]

[1] The department of Electrical and Computer Engineering, NJIT (grebel@njit.edu)
[2] Tuchman School of Management, NJIT
[3] The Center for Energy Efficiency, Resilience and Innovation, (CEERI) NJIT

**Abstract:** The energy market, and specifically the renewable sector carries volatility and risks, similar to the financial market. Here, we leverage on a well-established, return-risk approach, commonly used by equity portfolio-managers and apply it to energy resources. We visualize the relationship between the resources' costs and their risks in terms of efficient frontiers. We apply this analysis to publically available data for various US regions: Central, Eastern and Western coasts. Since risk management is contingent on costs, this approach sheds useful light in assessing dynamic pricing in modern electrical grids. By integrating geographical and temporal dimensions into our research, we aim at providing more nuanced and context-specific recommendations for energy resource allocation. This approach may help decision-makers in the renewable energy sector to make informed choices that account for regional variations, climatic conditions, and long-term performance trends.

**I. Introduction**

Commercial entities, such as, Commercial Real Estate (CRE) operate as profit-driven businesses, but their decision-making process is also influenced by social and environmental factors. Considerations of climate change when issuing government permits, highlight the broader impact of energy supply and demand. Other factors, which could be more international in kind, increase the risk factors as well, e.g., prices of oil and gas that are tied to local disputes and safe transportation. This study focuses on managing Distributed Energy Resources (DERs) within private and commercial entities that use diverse power resources; specifically, gas and diesel-powered generators, solar and wind farms, and energy storage. Similar to equity portfolios, these need to be balanced in order to optimize costs vs risks. By integrating financial technology models (FinTech) into the decision-making process, this paper aims to achieve a comprehensive risk assessment and management approach for power assets that incorporate sustainable resources. The proposed approach contribute to maintaining a balanced supply-demand relationship, system reliability, stability and power resource efficiency, least of which lead to a more rigorous approach to dynamic pricing.

Currently, there is reluctance to invest in, and manage DERs due to insufficient quantified uncertainties that lead to risk assessment. The finance, insurance, and actuarial sectors' risk management practices have not, yet, been fully incorporated within the management of the power grid. Here we seek to adapt risk assessment and management techniques used for stock portfolios and adopt them to energy grid management as a whole, or, as stand-alone commercial properties. It aims to provide valuable insights into effectively mitigating costs vs risks. Uncertainties in the operation of sustainable grid are due to individual and system-level interactions, such as weather and local supply and demand fluctuations. Similar to correlations that exist between stocks (e.g. the transportation sector), power grid assets are closely interconnected, as well (e.g., weather related wind and solar farms), while the energy commodity significantly influences the assets' value.

The efficiency frontier, also known as the efficient frontier, is a concept in portfolio management that illustrates the trade-off between risk and return for a given set of investment opportunities [1,2]. It is represented as a graph, or a curve that depicts the optimal portfolios that offers the highest expected return for a specific level of risk; conversely, it may be viewed as the lowest level of risk for a given target return. The efficiency frontier is derived through the process of portfolio optimization, which involves constructing portfolios that maximize returns while minimizing risk. The goal is to identify the most efficient portfolios that offer the best risk-return trade-offs. To construct the efficiency frontier, various asset allocations or combinations are considered. Each portfolio on the frontier represents a unique mix of assets with various expected returns and risk levels. These portfolios are generated by combining different proportions of assets with varying risk and return characteristics, such as stocks, bonds, or other financial instruments.

The efficiency frontier graphically illustrates the range of available portfolios and their risk-return profiles. Portfolios located on the frontier are considered efficient because they offer the maximum expected return for a given level of risk. Portfolios lying below the frontier are considered sub-optimal because they either offer lower returns for a given level of risk, or higher risk for a specific target return. The efficiency frontier is an essential tool for portfolio managers and investors; it helps them identify the optimal asset allocations based on their risk preferences and return objectives. By considering portfolios along the frontier, investors can make informed decisions about diversification and asset allocation to build portfolios that align with their risk tolerance and financial goals. In the following we treat the various grid's energy assets as entities that carry cost and risk and, thereby, need to be balanced and optimized.

We conducted an in-depth analysis of frontier efficiency graphs using publically available data. By studying these graphs, we aim to gain insights into the efficiency levels risks of various energy generation resources. As an example, we focused on three key energy assets: solar, wind, and biodiesel, the latter is included as a backup resource. These renewable energy sources play a crucial role in the transition towards a more sustainable and environmentally friendly power grid. By examining the output of energy generation from these resources with their underlined risks (e.g., intermittent supply, yet lower costs when available) we aimed to draw analogies with the patterns observed in frontier efficiency graphs.

In the context of literature, portfolios of physical products and financial instruments have been studied using finance/investment principles such as Modern Portfolio Theory (MPT) and Capital Asset Pricing Model (CAPM). For related literature, the reader is referred to [3,4], and references therein.

By analyzing the output of solar, wind, and biodiesel energy generation, we aimed to assess their efficiency levels, production capacities, and potential areas for improvement. This comparative analysis would provide valuable insights into the performance and viability of these resources. Moreover, by utilizing the framework of frontier efficiency graphs, we seek to establish a meaningful benchmark for evaluating the effectiveness and efficiency of each energy resource, thereby contributing to the overall understanding of their role in sustainable energy management.

The efficiency frontier represents the ideal balance between risk and expected return for a given portfolio of energy resources. By exploring various combinations of the resources, we aimed to identify the percentage allocation that yields the desired expected return while considering a certain level of risk. By leveraging solar, wind, and biodiesel in combination, we intended to determine the most effective and efficient mix that maximizes energy generation while minimizing potential risks. This approach enables us to assess the trade-offs between different energy sources and determine the optimal portfolio allocation that achieves the desired performance within acceptable risk parameters. By identifying the specific percentage allocation of solar, wind, and biodiesel resources, we aim to guide decision-making processes for energy managers and investors. This information allows them to make informed choices about resource allocation, considering both the potential returns therefore, determine pricing.

Ultimately, our objective is to uncover the optimal combination of solar, wind, and biodiesel resources that strikes a balance between risk and expected return (when a proper pricing is attached to the generation portion of it), and may contribute to the overall understanding of renewable energy management. Obviously, other assets may be incorporated into the analysis: gas generators, fuel cells, energy storage units to name a few. Such approach may facilitate the development of strategies that enhance efficiency, sustainability, and financial viability in the power generation sector.

Following our initial analysis, we proceeded with a more comprehensive investigation by comparing the efficiency and performance of the three combinations of solar, wind, and biodiesel across three distinct geographical locations (Eastern, central and western USA). This comparative analysis allowed us to assess how the optimal allocation of these resources may vary based on regional characteristics and environmental conditions. By examining multiple geographical locations, we aimed at capturing the diversity of energy resource availability, climate patterns, and market dynamics. This approach provides a more robust understanding of the effectiveness and adaptability of the different combinations in various contexts.

To ensure accuracy and reliability, we utilized data spanning three years to plot the efficiency frontier graphs and conducted further analysis. This longitudinal approach allowed us to capture seasonal variations, temporal trends, and overall performance patterns of the different combinations. By incorporating multiple years of data, we aimed to obtain a comprehensive view of the performance stability and consistency of the selected combinations. Through graph plotting and subsequent analysis, we aimed to uncover insights into the trade-offs between risk and expected return for each geographical location. This detailed examination enabled us to identify any significant variations in efficiency, risk profiles, and overall performance across different regions and time periods. By integrating geographical and temporal dimensions into our research, we aimed to provide more nuanced and context-specific recommendations for energy resource allocation. This approach helps decision-makers in the renewable energy sector to make informed choices that account for regional variations, climatic conditions, and long-term performance trends. Overall, our investigation expanded beyond a single combination of energy resources and incorporated geographical and temporal dimensions, allowing us to derive valuable insights into the efficiency frontier and its implications for renewable energy management.

## II. Data Collection and Analysis (Risk Assessment)

In modern portfolio theory, the efficient frontier (or portfolio frontier) is an investment portfolio which occupies the "efficient" parts of the risk–return spectrum [1,2]. Formally, it is the set of portfolios which satisfy the condition that no other portfolio exists with a higher expected return but with the same standard deviation of return (i.e., the risk). The efficient frontier was first formulated by Harry Markowitz in 1952[1].

---

[1] https://en.wikipedia.org/wiki/Efficient_frontier

In finance, the Sharpe ratio (also known as the Sharpe index, the Sharpe measure, and the reward-to-variability ratio), named after William F. Sharpe who coined the term in 1966. Sharpe ratio measures the performance of an investment such as a security or portfolio compared to a risk-free asset, after adjusting for its risk. It is defined as the difference between the returns of the investment and the risk-free return, divided by the standard deviation of the investment returns. It represents the additional amount of return that an investor receives per unit of increase in risk. The Sharpe ratio is defined as:

$$S_a = \frac{\mathrm{E}[r_a - r_f]}{\sigma_a}, \qquad (1)$$

where $r_a$ is the asset return, $r_f$ is the risk-free return (such as a U.S. Treasury security); $\mathrm{E}[r_a - r_f]$ is the expected value of the excess of the asset return over the benchmark return, and $\sigma_a$ is the standard deviation of the asset excess return.

**II.1 Analysis for one location – (Newark, NJ)**

To initiate the assessment of solar and wind energy resources, we collected relevant data using Newark's zip code as our starting point. We utilized a solar calculator tool provided by the National Renewable Energy Laboratory (NREL), which allowed us to make estimations of energy generation based on factors such as precise location coordinates, system size, and other relevant parameters. With the help of this solar calculator, we obtained valuable insights into the potential solar energy production for the Newark area for our required system specifications.

In addition to solar data, we obtained actual wind speed data from Weather Underground for the corresponding geographic location. This data allowed us to understand the wind resource potential in the area. By combining solar and wind data, we aimed to evaluate the complementary nature of these renewable energy sources and their potential for integration in commercial real estate projects. To complete the trifecta of energy resources, we incorporated biodiesel data from the Energy Information Administration (EIA). By integrating biodiesel as a third component, we aimed to explore the potential benefits and synergies between solar, wind, and biodiesel in the context of commercial real estate. Using efficiency frontier model, we first got all the values representing the different combinations of these three energy sources. Then, these values on the efficiency frontiers enable us to identify the most effective and efficient combinations that provide the highest expected return for the associated level of risk. By analyzing this frontier, we aim to propose optimal alternatives for the commercial real estate industry sector, allowing them to maximize the benefits derived from the integration of renewable energy sources. This analysis also aims to serve as a guideline for the decision-making processes, enabling stakeholders in the industry to leverage renewable energy resources effectively and maximize the benefits in terms of both financial returns and sustainability. All of these are depicted in Table 1 and Figure 1.

Table 1: summary of the various results

| Minimum Variance Portfolio | |
| --- | --- |
| Expected return for portfolio = E($r_p$) | 1.93% |
| (Risk) Std Dev. of portfolio = $\sigma_p$ = $(w\Sigma w^T)^{1/2}$ | 12.37% |

| Minimum variance efficiency portfolio (Maximize Sharpe ratio ) | |
| --- | --- |
| Expected return for portfolio = E($r_p$) | 2.44% |
| (Risk) Std Dev. of portfolio = $\sigma_p$ = $(w\Sigma w^T)^{1/2}$ | 13.91% |

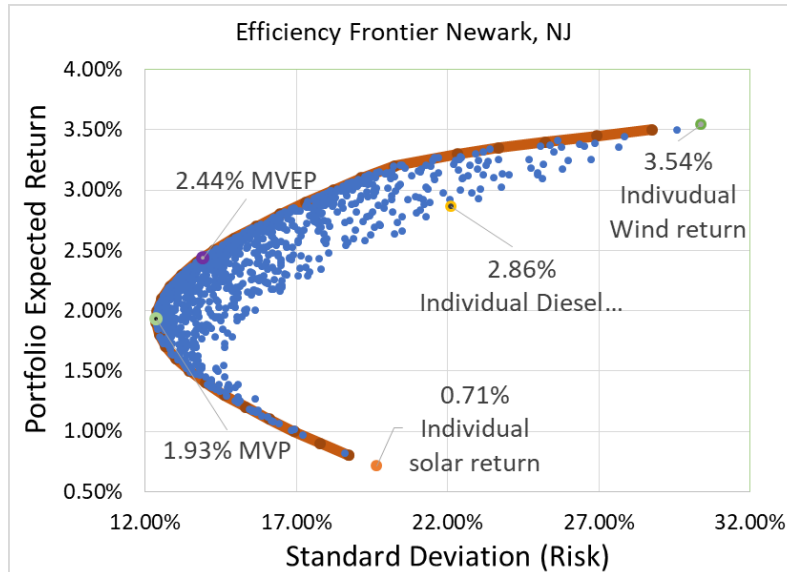

Figure 1. Portfolio expected Return vs Standard deviation (Risk, in percentile points) for a yearly consumption in the area of Newark, NJ.

In the context of the efficiency frontier, the Minimum Variance Portfolio (MVP) and Minimum Variance Efficiency Portfolio (MVEP) are key concepts that relate to the trade-off between risk and return in portfolio optimization. The MVP refers to a portfolio that exhibits the lowest possible level of variance or volatility among all possible portfolios with a given level of expected return. It represents the portfolio that minimizes the overall risk while achieving a specific target return. The MVP is an essential point on the efficiency frontier, as it represents the portfolio with the lowest risk for a particular level of expected return.

On the other hand, the MVEP is the portfolio on the efficiency frontier that offers the highest level of risk-adjusted return, often measured by the Sharpe ratio. It represents the optimal combination of assets that maximizes the expected return while minimizing the risk. The MVEP is considered the most efficient point on the efficiency frontier, as it provides the highest level of return for a given level of risk or the lowest level of risk for a specific target return.

Both the MVP and MVEP are crucial in portfolio optimization as they represent key points along the efficiency frontier, illustrating the relationship between risk and return. The MVP highlights the portfolio with the minimum risk for a given level of return, while the MVEP showcases the portfolio with the best risk-adjusted return among all possible combinations. These concepts aid investors and portfolio managers in constructing optimal portfolios that align with their risk preferences and return objectives.

**II.2 Comparing data for one location (Newark, NJ) over 3 years**

The second phase of our study involved analyzing the data collected over a span of three years for a specific location, Newark, New Jersey. In this study, the data collection process incorporates the inclusion of derivative values. Unlike the first part of the study where the exact solar and wind energy was calculated based on specific system sizes, the analysis in this phase relies on actual solar radiance data provided by the National Renewable Energy Laboratory (NREL) and the recorded wind speeds for the corresponding location over the duration of the study. Incorporating derivative values in the analysis does not impact the final efficiency frontier values since all the calculations are based on relative return.

By examining the efficiency frontier graphs for each year, we made noteworthy observations regarding the portfolio combinations and risk variations. But first it is important to understand the meaning of correlation in the context of the efficiency frontier model. Correlation refers to the statistical measure that quantifies the degree of association or relationship between the returns or performance of different portfolio assets. It indicates the extent to which the values of the assets move together or diverge from each other. Correlation is typically measured on a scale from -1 to 1. A correlation coefficient of +1 represents a perfect positive correlation, indicating that the assets move in perfect tandem, while a correlation coefficient of -1 represents a perfect negative correlation, indicating that the assets move in opposite directions. A correlation coefficient of 0 suggests no correlation, indicating that the assets' movements are independent of each other.

When constructing the efficiency frontier, the correlation among portfolio assets is a crucial factor to consider (Table 2). Positive correlations among assets imply that they tend to move together, which can lead to lower diversification benefits and potentially increase the overall portfolio risk. Conversely, negative correlations indicate that assets move in opposite directions, which can provide diversification benefits and potentially reduce the portfolio risk. By analyzing the correlations among the portfolio assets, investors and portfolio managers can gain insights into how different assets interact with each other and affect the overall risk and return of the portfolio. It helps them identify asset combinations that offer complementary risk and return characteristics, enabling the construction of more efficient portfolios along the efficiency frontier.

Table 2. Relative correlation factors: Correlation for year 2021

| Correlation | Solar | Wind | Diesel |
|---|---|---|---|
| **Solar** | 1 | 0.11001482 | 0.46258328 |
| **Wind** | 0.110014817 | 1 | -0.2612327 |
| **Diesel** | 0.462583279 | -0.2612327 | 1 |

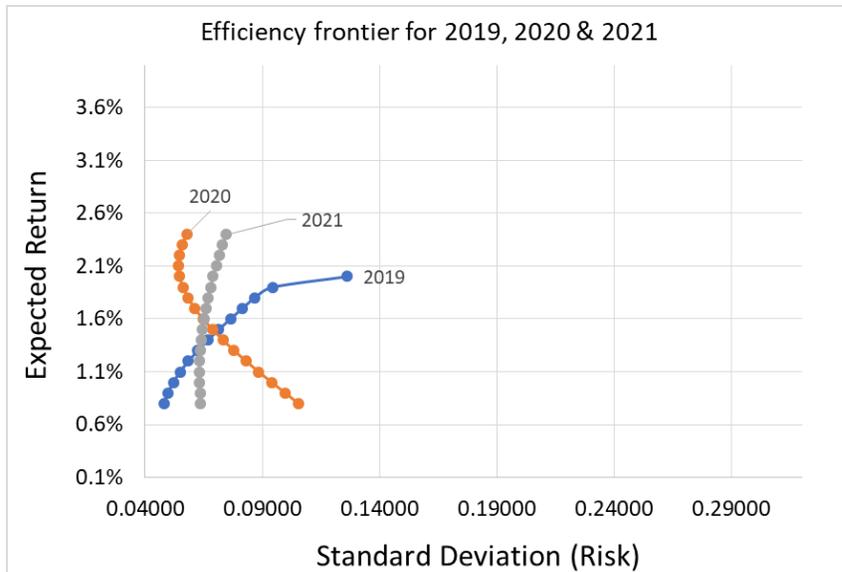
Figure 2. Efficiency frontiers for a 3-year time span.

In Fig. 2 we present the efficiency frontiers over the three-year span. It becomes apparent that the frontier for the year 2021 displayed the most efficient portfolio combinations with minimal variations in risk. This finding can be attributed to the positive correlation observed between solar and wind energy sources during that year. The positive correlation implies that when solar energy generation increased, wind energy generation also exhibited a corresponding increase, leading to a more harmonious and complementary relationship between these two energy sources. As a result, portfolios constructed in 2021 were able to achieve higher returns with relatively lower levels of risk compared to the other two years.

In contrast, the efficiency frontiers for the remaining two years demonstrated a negative correlation between solar and wind energy sources. This negative correlation indicates that during those years, solar and wind energy generation did not align or exhibit a consistent relationship. Consequently, portfolios constructed in those years faced greater risk and exhibited higher variations in risk compared to the more positively correlated year of 2021. This analysis highlights the significance of understanding the correlations between different energy sources when constructing efficient portfolios. By recognizing the interplay between solar and wind energy sources and their impact on portfolio risk, stakeholders in the commercial real estate industry can make decisions about energy resource allocation and portfolio optimization strategies. Among the years analyzed, the year 2020 stands out with the highest expected return of 2.4% and the least amount of risk measured at 0.058. It represents the potential profitability of the investment and implies a greater potential for generating profits from the selected assets. The assets distribution contributing to this value is 2.93% solar, 20.94% wind and 76.14% biodiesel. By considering the risk-return profiles of different years and the distribution of assets over these combinations it is easier to decide how one wants to invest in the assets covering the overall portfolio to get the desired return while keeping risk at minimum.

**II.3 Comparing three geographical locations (Eastern, Central & Western)**

In this specific segment of the study, we conducted a comparison of the efficiency frontiers for three distinct geographical locations: Newark, New Jersey (Eastern USA), Lincoln, Nebraska (Central USA), and Los Angeles, California (Western USA) as shown in Figure 3. By analyzing the efficiency frontier graphs, we were able to gain insights into the risk-return profiles of various geographical regions.

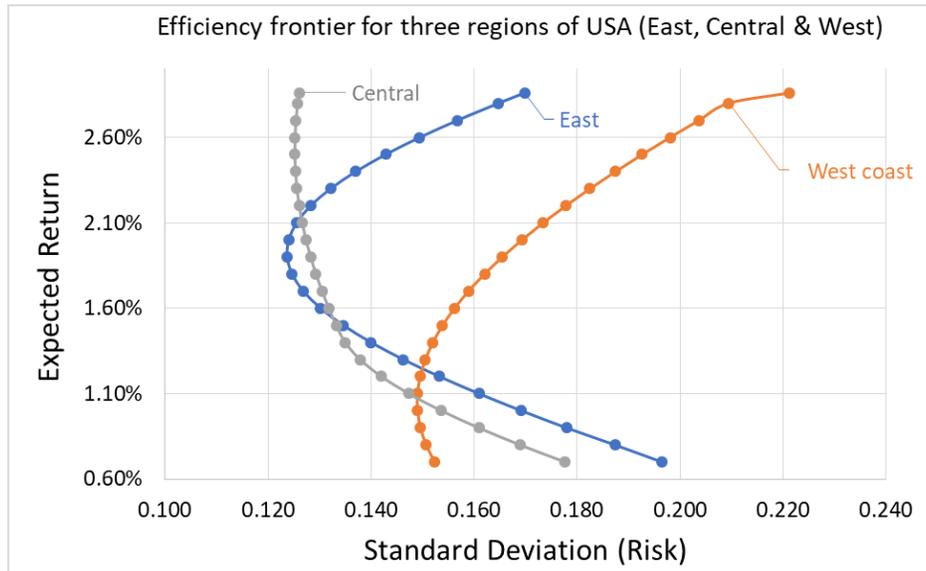

Figure 3. Efficiency frontiers for three geographical regions

As depicted in the graph, the central region, represented by Lincoln, Nebraska, tends to exhibit the lowest levels of risk across a wide range of expected return values. This suggests that portfolios constructed in the central region have a more favorable risk profile compared to the other regions. Conversely, the western region, represented by Los Angeles, California, shows higher levels of risk for the corresponding expected return values. This indicates that portfolios in the western region tend to be riskier in nature. In contrast, we observed the lowest risk value of 0.124 for an expected return of 1.9% in the east coast region, specifically Newark, New Jersey. This indicates that portfolios in the east coast region have the most favorable risk-return trade-off at this return level. The portfolio assets, in our case energy sources, were distributed as follows for this specific return: 50.54% solar, 18.62% wind, and 30.84% biodiesel.

**III. Discussions and Conclusion**

In this paper we propose to treat distributed energy assets as a portfolio of equities and analyze the results for optimal return vs risks. We compared short term analysis (one year) and longer term analysis (3 years) to show that similarly to stocks, one has to frequently assess the expected return vs risk. Visually, this has been done by use of efficiency frontiers curves. Variations in expected return/risk profiles were observed not only over time but over various US regions. Since the costs (development and maintenance) for these assets are known, the point of fully covering costs and turning into profit may be factored in the expected return/risk analysis. This type of analysis offers valuable insights for the decision-makers in commercial real estate and grid management alike. It also helps making informed choices when allocating energy resources, help establish dynamic pricing scheduling and overall, helps achieving a balanced risk-return objectives.

**Acknowledgement:** this project was funded in part by Paul Profeta Real Estate Center - Faculty Seed Grant 2022, NJIT and the Port America.